\documentclass{jfm}
\usepackage{graphicx}
\usepackage{color}
\usepackage[dvipsnames]{xcolor}

\usepackage{siunitx}
\usepackage{subfig}

\shorttitle{Finite-sized rigid spheres in turbulent Taylor-Couette flow}
\shortauthor{D. Bakhuis, R. A. Verschoof, V. Mathai,  S. G. Huisman, D. Lohse and  C. Sun}

\title{Finite-sized rigid spheres in turbulent Taylor-Couette flow: Effect on the overall drag}

\author[D. Bakhuis, R. A. Verschoof, V. Mathai, S. G. Huisman, D. Lohse
and C. Sun]%
{Dennis Bakhuis$^1$%
,\ns
Ruben A. Verschoof$^1$%
,\ns
Varghese Mathai$^1$,\ns\break%
Sander G. Huisman$^{1}$,\ns%
Detlef Lohse$^{1,2}$,\ns and\ns%
Chao Sun$^{3,1}$}

\affiliation{%
$^1$Physics of Fluids Group, Max Planck UT Center for Complex Fluid Dynamics,\break
MESA+ Institute and J.M. Burgers Centre for Fluid Dynamics,\break
University of Twente, P.O. Box 217, 7500 AE Enschede, The Netherlands\\
$^2$Max Planck Institute for Dynamics and Self-Organization, 37077
G\"ottingen, Germany\\
$^3$Center for Combustion Energy and Department of Thermal Engineering,
Tsinghua University, 100084 Beijing, China}

\newcommand\Nus{\text{Nu}}  
\newcommand\Tay{\text{Ta}}  
\newcommand\rey{\text{Re}}  

\usepackage{tikz}
\usetikzlibrary{plotmarks}

\newcommand\marksymbol[2]{\tikz[#2,scale=1.2]\pgfuseplotmark{#1};}

\newcommand{\red}[1]{\textcolor{black}{#1}}

\begin{document}

\maketitle

\begin{abstract}
We report on the modification of drag by neutrally buoyant spherical
finite-sized particles in highly turbulent Taylor-Couette (TC) flow.  These
particles are used to disentangle the effects of size, deformability, and
volume fraction on the drag, and are contrasted with the drag in bubbly TC
flow.  From global torque measurements we find that rigid spheres hardly
decrease or increase the torque needed to drive the system.

The size of the particles under investigation have a marginal effect on the
drag, with smaller diameter particles showing only slightly lower drag.
Increasing the particle volume fraction shows a net drag increase, however
this increase is much smaller than can be explained by the increase in
apparent viscosity due to the particles.  The increase in drag for increasing
particle volume fraction is corroborated by performing laser Doppler
anemometry where we find that the turbulent velocity fluctuations also
increase with increasing volume fraction. In contrast with rigid spheres, for
bubbles the effective drag reduction also increases with increasing Reynolds
number. Bubbles are also much more effective in reducing the overall drag. 
\end{abstract}


\section{Introduction}\label{sec:introduction}
Flows in nature and industry are generally turbulent, and often these flows
carry bubbles, drops, or particles of various shapes, sizes, and densities.
Examples include sediment-laden rivers, gas-liquid reactors, volcanic
eruptions, plankton in the oceans, pollutants in the atmosphere, and air
bubbles in the ocean mixing layer~\citep{Toschi2009}.
Particle-laden flows may be characterized in terms of particle density
$\rho_p$, particle diameter $d_p$, volume fraction $\alpha$, and Reynolds
number Re of the flow. When $d_p$ is small (compared to the dissipative length
scale $\eta_K$) and $\alpha$ low ($< 10^{-3}$), the system may be modelled
using a point particle approximation with two-way
coupling~\citep{Elghobashi1994,Mazzitelli2003,Mathai2016}.
With recent advances in computing, fully resolved simulations of
particle-laden flows have also become feasible. \cite{Uhlmann2008} conducted
one of the first numerical simulations of finite-sized rigid spheres in a
vertical particle-laden channel flow. They observed a modification of the mean
velocity profile and turbulence modulation due to the presence of particles. A
number of studies followed, which employed immersed
boundary~\citep{Peskin2002,Cisse2013}, Physalis~\citep{Naso2010, Wang2017b},
and front-tracking methods~\citep{Unverdi1992,Roghair2011,Tagawa2013} to treat
rigid particles and deformable bubbles, respectively, in channel and pipe flow
geometries~\citep{Pan1996,Lu2005,Uhlmann2008,Dabiri2013,Kidanemariam2013,
Lashgari2014,Picano2015,Costa2016}. 
Flows with dispersed particles, drops, and bubbles can, under the right
conditions, reduce skin friction and result in significant energetic (and
therefore financial) savings. In industrial settings this is already achieved
using polymeric additives which disrupt the self-sustaining cycle of wall
turbulence and dampen the quasi-streamwise vortices
\citep{White2008,Procaccia2008}. Polymeric additives are impractical for
maritime applications, and therefore gas bubbles are used with varying success
rates \citep{Ceccio2010,Murai2014}. Local measurements in bubbly flows are
non-trivial and the key parameters and their optimum values are still unknown.
For example, it is impossible to fix the bubble size in experiments and
therefore to isolate the effect of bubble size. Various studies hinted that
drag reduction can also be achieved using spherical particles
\citep{Zhao2010}, also by using very large particles in a turbulent von
K\'arm\'an flow \citep{Cisse2015}. In this latter study a tremendous decrease
in turbulent kinetic energy (TKE) was observed. A similar, but less intense,
decrease in TKE was also seen by \cite{Bellani2012b} using a very low particle
volume fraction. By using solid particles it is possible to isolate the size
effect on drag reduction and even though rigid particles are fundamentally
different from bubbles, this can give additional insight into the mechanism of
bubbly drag reduction. \cite{Machicoane2016} already showed that the particle
dynamics are highly influenced by the diameter of the particle. This might or
might not have a direct influence on the global drag of the system and has
never been studied.
Whether and when solid particles increase or decrease the drag in a flow is
yet not fully understood and two lines of thought exist. On one side, it is
hypothesized that solid particles {\it decrease} the overall drag as
they damp turbulent fluctuations \citep{Zhao2010,Poelma2007}. On the
other side, one could expect that solid particles {\it increase} drag as they
shed vortices, which must be dissipated. In addition, they also increase the
apparent viscosity. A common way to quantify this is the so called `Einstein
relation' (\cite{Einstein1906}:
\begin{equation}
\nu_\alpha = \nu\left(1 + \frac{5}{2}\alpha \right),
\label{eq:einstein}
\end{equation}
where $\nu$ is the viscosity of the continuous phase. This compensation
is valid for the small $\alpha$ values used in this manuscript
\citep{Stickel2005}. Direct measurements of drag in flows with solid
particles are scarce, and the debate on under what condition they either
enhance or decrease the friction has not yet been settled.
Particles and bubbles may show collective effects (clustering) and
    experiments have revealed that this has significant influence on the flow
    properties~\citep{Liu1993, Kulick1994, Muste1997, So2002, Fujiwara2004,
    vandenBerg2005, vandenBerg2007, Shawkat2008, Calzavarini2008, Colin2012,
vanGils2013, Maryami2014, Mathai2015, Almeras2017,mathai2018enhanced}. In
general, the Stokes number is used to predict this clustering behaviour, but
for neutrally buoyant particles this is found to be insufficient
\citep{Bragg2015,Fiabane2012}. In addition, the position of the particles (or
the particles clusters) is likely to have a large influence on the skin
friction. 
In DNS at low Reynolds numbers, \cite{Kazerooni2017} found that the particle
distribution is mainly governed by the bulk Reynolds number.

In order to study the effects of particles on turbulence it is convenient to
use a closed setup where one can relate global and local quantities directly
through rigorous mathematical relations. In this manuscript the Taylor-Couette
(TC) geometry~\citep{Grossmann2016}---the flow between two concentric rotating
cylinders---is employed as this is a closed setup with global balances. The
driving of the Taylor-Couette geometry can be described using the Reynolds
number based on the inner cylinder (IC): $\rey_i = u_i d / \nu$, where $u_i =
\omega_i r_i$ is the azimuthal velocity at the surface of the IC, $\omega_i$
the angular velocity of the IC, $d = r_o - r_i$ the gap between the cylinders,
$\nu$ the kinematic viscosity, and $r_i$($r_o$) the radius of the inner(outer)
cylinder. The geometry of Taylor-Couette flow is characterized by two
parameters: the radius ratio $\eta=r_i/r_o$ and the aspect ratio $\Gamma =
L/d$, where $L$ is the height of the cylinders. The response parameter of the
system is the torque, $\tau$, required to maintain constant rotation speed of
the inner cylinder. It was mathematically shown that in Taylor-Couette flow
the angular velocity flux defined as $J^\omega = r^3 \left(  \left
\langle u_r \omega \right \rangle_{A,t} - \nu \frac{\partial} {\partial r}
\left \langle \omega \right \rangle_{A,t} \right)$, where the subscript $A,t$
denotes averaging over a cylindrical surface and time, is a radially
conserved quantity (\cite*{Eckhardt2007} (EGL)). One can, in analogy to
Rayleigh-B\'enard convection, normalize this flux and define a Nusselt number
based on the flux of the angular velocity:
\begin{equation}
  \Nus_\omega = \frac{J^\omega}{J^\omega_{\text{lam}}} = \frac{\tau}{2 \pi L \rho J^\omega_{\text{lam}}},
\end{equation}
where $J^\omega_{\text{lam}} = 2 \nu r^2_i r^2_o \left( \omega_i - \omega_o \right)/\left( r^2_o - r^2_i \right)$ is the angular velocity flux for laminar, purely azimuthal flow and $\omega_o$ is the angular velocity of the outer cylinder. In this spirit the driving is expressed in terms of the Taylor number:
\begin{equation}
  \Tay = \frac{1}{4} \sigma d^2 \left(r_i + r_o \right)^2 \left( \omega_i - \omega_o \right)^2 \nu^{-2}.
\end{equation}
Here $\sigma = \left( \left( 1 + \eta \right) / \left(2 \sqrt{\eta} \right) \right)^4\;\approx\;1.057$ is a geometric parameter (``geometric Prandtl number''), in analogy to the Prandtl number in Rayleigh-B\'enard convection. In the presented work, where only the inner cylinder is rotated and the outer cylinder is kept stationary, we can relate $\Tay$ to the Reynolds number of the inner cylinder by
\begin{equation}
\rey_i = \frac{r_i \omega_i d}{\nu} = \frac{8\eta^2}{(1+\eta)^{3}} \sqrt{\Tay}.
\end{equation}
\noindent The scaling of the dimensionless angular velocity flux (torque) with
the Taylor (Reynolds) number has been analysed extensively, see e.g.
\cite{Lathrop1992,Lewis1999,Paoletti2011,vanGils2011,Ostilla-Monico2013} and
the review articles by \cite{Fardin2014} and \cite{Grossmann2016}, and the
different regimes are well understood. In the current Taylor number regime it
is known that $\Nus_\omega \propto \Tay^{0.4}$. Because this response is
well known, it can be exploited to study the influence of immersed bubbles and
particles
~\citep{vandenBerg2005,vandenBerg2007,vanGils2013,Maryami2014,Verschoof2016}
on the drag needed to sustain constant rotational velocity of the inner
cylinder. 

In this paper we will use the TC geometry to study the effect of neutrally
buoyant rigid spherical particles on the drag. We study the effects of varying
the particle size $d_p$, the volume fraction $\alpha$, the density ratio
$\phi$, and the flow Reynolds number $\text{Re}$ on the global torque (drag)
of the Taylor-Couette flow. The drag reduction is expressed as $\text{DR} =
\left(1 - \Nus_\omega(\alpha) / \Nus_\omega(\alpha=0) \right)$ and as we are
interested in the net drag reduction, it is \emph{not} compensated for
increased viscosity effects using correction models, such as the Einstein
relation.

The manuscript is organized as follows. Section \ref{sec:experimentalSetup} presents the experimental setup. In section \ref{sec:results} we discuss the results. The findings are summarized and an outlook for future work is given in the last section.

\section{Experimental setup}\label{sec:experimentalSetup}
\begin{figure}
\centering
\includegraphics[height=7cm]{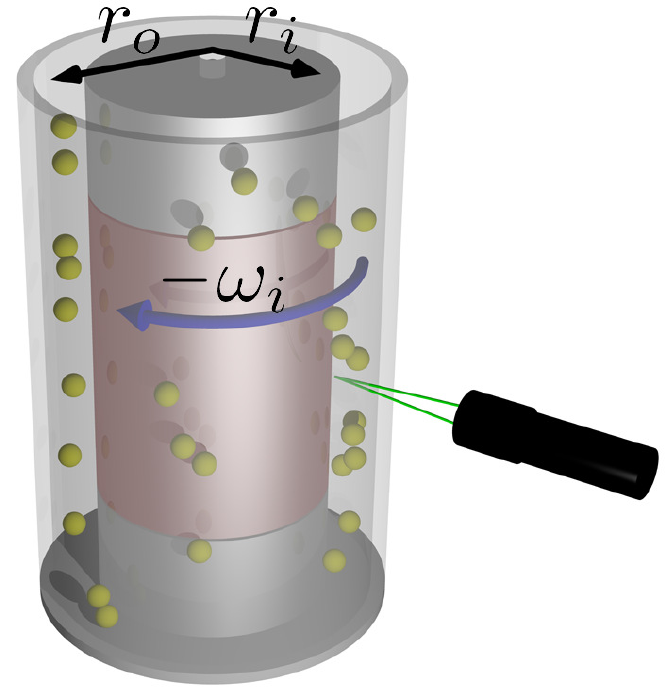}
\caption{%
Schematic of the Taylor-Couette setup. Two concentric cylinders of radii
$r_{i,o}$ with a working fluid in between. Particles are not to scale. The
inner cylinder rotates with angular velocity $\omega_i$, while the outer
cylinder is kept at rest. We measure the torque on the middle section
(highlighted). The laser Doppler anemometry (LDA) probe is positioned at mid
height to measure the azimuthal velocity at mid gap.}
\label{fig:setup}
\end{figure}

The experiments were conducted in the Twente Turbulent Taylor-Couette
(T${}^3$C) facility (\cite{vanGils2011}). A schematic of the setup is shown in
Figure \ref{fig:setup}. In this setup, the flow is confined between two
concentric cylinders, which rotate independently. The top and bottom plates
are attached to the outer cylinder. The radius of the inner cylinder (IC) is
$r_i = \SI{0.200}{\metre}$ and the radius of the outer cylinder (OC) is $r_o =
\SI{0.2794}{\metre}$, resulting in a gap width of $d = r_o - r_i =
\SI{0.0794}{\metre}$ and a radius ratio of $\eta = r_i/r_o = 0.716$. The IC
has a total height of $L = \SI{0.927}{\metre}$ resulting in an aspect ratio of
$L / d = 11.7$. The IC is segmented axially in three parts. To minimize the
effect of the stationary end plates, the torque is measured only over the
middle section of the IC with height $L_{\text{mid}}/L = 0.58$, away from
the end plates. A hollow reaction torque sensor made by Honeywell is used to
measure the torque which has an error of roughly 1\% for the largest torques
we measured. Between the middle section and the top and bottom section of the
inner cylinder is a gap of 2mm.

The IC can be rotated up to $f_i = \omega_i/(2\pi) =
\SI{20}{\hertz}$. In these experiments only the IC is rotated and the OC is
kept at rest. The system holds a volume of $V = \SI{111}{\litre}$ of working
fluid, which is a solution of glycerol ($\rho =
\SI{1260}{\kilo\gram/\metre\cubed}$) and water. To tune the density of the
working fluid, the amount of glycerol was varied between \SI{0}{\percent} and
\SI{40}{\percent} resulting in particles being marginally heavy,
neutrally buoyant, or marginally light.

The system is thermally controlled by cooling the top and bottom plates of the
setup. The temperature was kept at $T = \SI{20(1)}{\celsius}$ for all the
experiments, with a maximum spatial temperature difference of
$\SI{0.2}{\kelvin}$ within the setup, and we account for the density and
viscosity changes of water and glycerol \citep{Glycerine1963}.

Rigid polystyrene spherical particles (\textit{RGPballs S.r.l.}) were used in
the experiments, these particles have a density close to that of water
(940\,--\SI{1040}{\kilo\gram/\metre\cubed}). We chose particles with diameters
$d_p = 1.5$, 4.0, and $\SI{8.0}{\milli\metre}$. To our disposal are:
\SI{2.22}{\litre} of \SI{1.5}{\milli\metre} diameter particles,
\SI{2.22}{\litre} of \SI{4}{\milli\metre} diameter particles, and
\SI{6.66}{\litre} of \SI{8}{\milli\metre} diameter particles, resulting in
maximum volume fractions of \SI{2}{\percent}, \SI{2}{\percent}, and
\SI{6}{\percent}, respectively. The particles are found to be nearly
mono-disperse (\SI{99.9}{\percent} of the particles are within \SI{\pm
0.1}{\milli\metre} of their target diameter). Due to the fabrication process,
small air bubbles are sometimes entrapped within the particles. This results
in a slight heterogeneous density distribution of the particles. After
measuring the density distribution for each diameter, we calculated the
average for all batches, which is $\rho_p = \SI{1036 \pm
5}{\kilo\gram/\metre\cubed}$. By adding glycerol to water we match this value
in order to have neutrally buoyant particles.

Using a laser Doppler anemometry (LDA) system (BSA F80, Dantec Dynamics) we
captured the azimuthal velocity at mid-height and mid-gap of the system (see
figure \ref{fig:setup}) and we performed a radial scan at mid-height.
The flow was seeded with \SI{5}{\micro\metre} diameter polyamide particles
(PSP-5, Dantec Dynamics). Because of the curved surface of the outer cylinder
(OC), the beams of the LDA get refracted in a non-trivial manner, which was
corrected for using a ray-tracing technique described in \cite{Huisman2012}. 

Obviously, LDA measurements in a multi-phase flow are more difficult to set up
than for single phase flows, as the The method relies on the reflection of
light from tiny tracer particles passing through a measurement volume
($\SI{0.07}{\milli\metre} \times \SI{0.07}{\milli\metre} \times
\SI{0.3}{\milli\metre}$).  Once we add a second type of relatively large
particles to the flow, this will affect the LDA measurements, mostly by
blocking the optical path, resulting in lower acquisition rates. These large
particles will also move through the measurement volume, but as these
particles are at least 300 times larger than the tracers and thus much larger
than the fringe pattern (fringe spacing $d_f=\SI{3.4}{\micro\metre}$), the
reflected light is substantially different from a \emph{regular} Doppler burst
and does not result in a measured value. The minimal signal-to-noise ratio for
accepting a Doppler burst was set to 4. As a post-processing step the
velocities were corrected for the velocity bias by using the transit time of
the tracer particle.

\section{Results}\label{sec:results}

\subsection{Effect of particle size}\label{subsec:sizeEffect}
\begin{figure}
\centering%
\subfloat{%
  \includegraphics{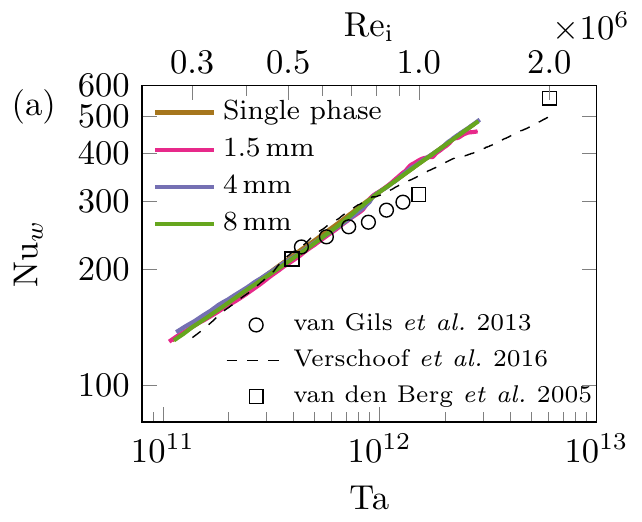}%
  \label{fig:TaNuwSize}
}%
\subfloat{%
  \includegraphics{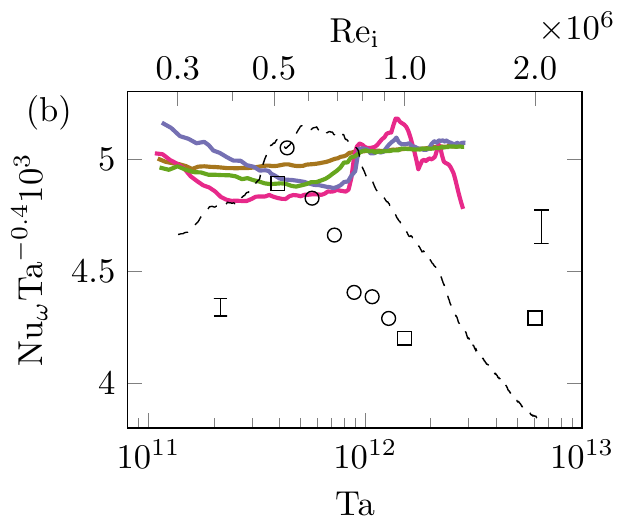}
  \label{fig:TaNuwTaSize}
}%
\caption{%
(a) $\Nus_\omega (\Tay)$ for \SI{2}{\percent} particle volume fraction with particle diameters of \SI{1.5}{\milli\metre}, \SI{4.0}{\milli\metre}, and \SI{8.0}{\milli\metre}, and for comparison the single phase case. Data from comparable bubbly drag reduction studies are plotted using black markers. (b) Same data, but now as compensated plot $\Nus_\omega / \Tay^{0.40}$ as function of $\Tay$.
\red{
The error bar indicates the maximum deviation for repeated measurements from all measurements combined (coloured curves) which is less than \SI{1}{\percent}. At $\Tay \geq \num{2e12}$, the \SI{1.5}{\milli\metre} particles show an increased uncertainty of \SI{1.7}{\percent}, which is indicated by the right error bar.}
}
\label{fig:dragReductionSize}
\end{figure}

First we study the effect of changing the particle diameter on the torque of the system. In these experiments, we kept the particle volume fraction fixed at \SI{2}{\percent} and the density of the working fluid, $\rho_f$, at \SI{1036}{\kilogram/\metre \cubed}, for which the particles are neutrally buoyant. The results of these measurements are presented as $\Nus_\omega(\Tay)$ in figure \ref{fig:TaNuwSize}. Our curves are practically overlapping, suggesting that the difference in drag between the different particle sizes is only marginal. We compare these with the bubbly drag reduction data at similar conditions (hollow symbols) from \cite{Verschoof2016,vanGils2013,vandenBerg2005}. At low $\Tay$ the symbols overlap with our data. However, at larger $\Tay$, the bubbly flow data shows much lower torque (drag) than the particle-laden cases. As we are in the ultimate regime of turbulence where $\Nus_\omega$ effectively \red{scales} as $\Nus_\omega \propto \Tay^{0.4}$~\citep{Huisman2012,Ostilla-Monico2013}, we compensate the data with $\Tay^{0.40}$ in figure \ref{fig:TaNuwTaSize} to 
\red{emphasize}
the differences between the datasets. For the single phase case, this yields a clear plateau. For the particle-laden cases, the lowest drag corresponds to the smallest particle size. The reduction is however quite small ($<\SI{3}{\percent}$). The compensated plots also reveal a sudden increase in drag at a critical Taylor number $\Tay^* =\num{0.8e12}$. The jump is more distinct for the smaller particles, and might suggest a reorganisation of the flow (\cite{Huisman2014}). Beyond $\Tay^*$, the drag reduction is negligible for the larger particles (\SI{4}{\milli \metre} and \SI{8}{\milli \metre} spheres). However, for the \SI{1.5}{\milli\metre} particles, the drag reduction seems to increase, and was found to be very repeatable in experiments. Interestingly, the size of these particles is comparable to that of the air bubbles in \cite{vanGils2013}.  This might suggest that for smaller size particles at larger $\Tay$, one could expect drag reduction. At the increased viscosity of the suspension, a maximum $\Tay \approx 3 \times 10^{12}$ could be reached in our experiments.
\red{
We have performed an uncertainty analysis by repeating the measurements
for the single phase, and for the cases with \SI{8}{\milli\metre} and \SI{1.5}{\milli\metre} particles multiple times and calculating the maximum deviation from the ensemble average.
The left error bar indicates the maximum deviation for all measurements combined and is $\approx \SI{1}{\percent}$.
For $\Tay \geq \num{2e12}$, we see an increase in uncertainty of \SI{1.7}{\percent} (shown by the right error bar in figure \ref{fig:TaNuwTaSize}), which is only caused by the \SI{1.5}{\milli\metre} particles. These tiny particles can accumulate in the \SI{2}{\milli\metre} gap between the cylinder segments and thereby increase the uncertainty. Above $\Tay \geq \num{2e12}$, both, the \SI{8}{\milli\metre} and \SI{4}{\milli\metre} particles, show a maximum deviation below 0.25\%.}

Below $\Tay^*$, the drag reduction due to spherical particles appears to be similar to bubbly drag reduction~\citep{vanGils2013}. However, in the lower $\Tay$ regime, the bubble distribution was highly non-uniform due to buoyancy of the bubbles~\citep{vandenBerg2005,vanGils2013,Verschoof2016}. Therefore, the volume fractions reported were only the global values, and the torque measurements were for the mid-sections of their setups. What is evident from the above comparisons is that in the high $\Tay$ regime, air bubbles drastically reduce the drag, reaching far beyond the drag modification by rigid spheres.

\subsection{Effect of particle volume fraction}\label{subsec:volumeFraction}
\begin{figure}
\centering%
\subfloat{%
  \includegraphics{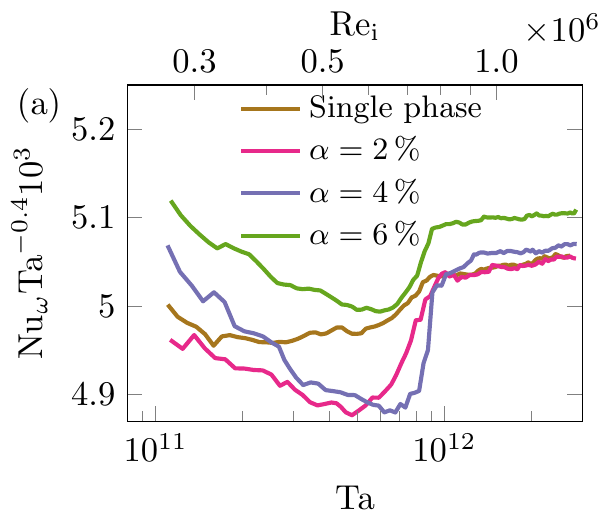}%
  \label{fig:TaNuwTaVF}
}%
\subfloat{%
  \includegraphics{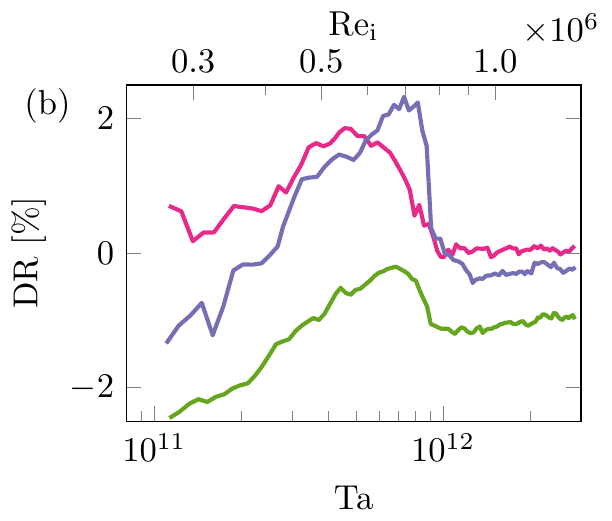}
  \label{fig:TadrVF}
}%
\caption{%
(a) $\Nus_\omega (\Tay)$, compensated by $\Tay^{0.4}$, for
\SI{8}{\milli\metre} particles with various particle volume fractions and for comparison the
single phase case.
\red{(b) Drag reduction, defined as $\text{DR} = \left(1 - \Nus_\omega(\alpha) / \Nus_\omega(\alpha=0) \right)$, plotted against $\Tay$.}
}
\label{fig:dragReductionVF}
\end{figure}

The next step is to investigate the effect of the particle volume fraction on
the torque. For the \SI{8}{\milli\metre} particles, we have the ability to increase the particle volume fraction up to \SI{6}{\percent}. This was done in steps of \SI{2}{\percent}, and the results are plotted in compensated form in figure \ref{fig:TaNuwTaVF}. The normalised torque increases with the volume fraction of particles. The \SI{6}{\percent} case shows the \red{largest} drag. Figure \ref{fig:TadrVF} shows the same data in terms of drag reduction as function of $\Tay$. A \SI{2}{\percent} volume fraction of particles gives the highest drag reduction. With increasing $\alpha$ the drag reduction decreases. These measurements are in contrast with the findings for bubbly drag reduction~\citep{vanGils2013}, for which the net drag decreases with increasing gas volume fraction.
\red{%
A common explanation for the increase of drag in a particle-laden flow is the larger apparent viscosity.
If we would calculate the apparent viscosity for our case with the Einstein relation (equation \ref{eq:einstein}) for $\alpha=\SI{6}{\percent}$, the drag increase would be \SI{15}{\percent}, as compared to the pure working fluid.
Including this effect in our drag reduction calculation would result in reductions of the same order.
However, when comparing the drag with or without particles, the \emph{net} drag reduction is practically zero.
This result is different from the work of \cite{Picano2015} in turbulent channel flow where they found that the drag increased more than the increase of the viscosity.
}

\begin{figure}
\centering%
\subfloat{%
  \includegraphics{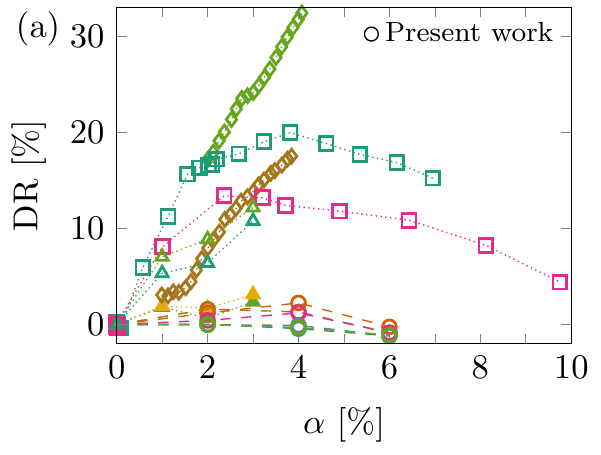}%
  \label{fig:alphaDrAll}
}%
\subfloat{%
  \includegraphics{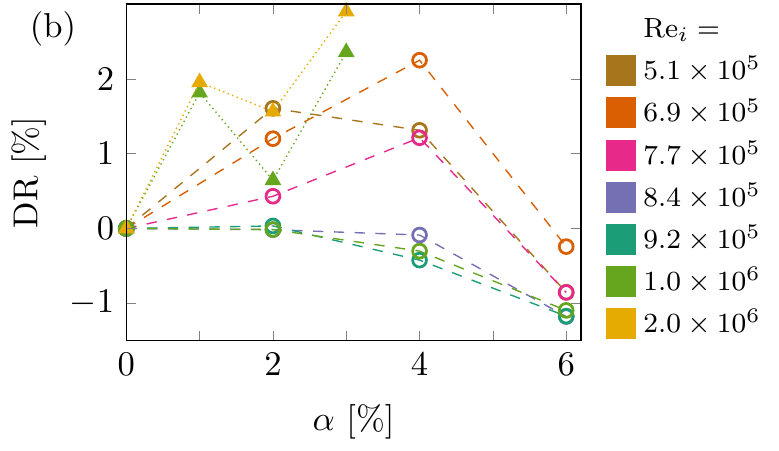}%
  \label{fig:alphaDrZoom}
}%
\caption[]{(a) Drag reduction as function of particle volume fraction from \marksymbol{o}{black}\,: \red{$d_p=\SI{8}{\milli\metre}$ particles from present work} compared to similar gas
volume fractions from \marksymbol{square}{black}\,: \cite{vandenBerg2005}, \marksymbol{diamond}{black}\,: \cite{vanGils2013}, and \marksymbol{triangle}{black}\,: \cite{Verschoof2016}. Symbols indicate the different studies while colours differentiate between the Reynolds numbers. 
\red{
The current work has DR defined as $\left(1 - \Nus_\omega(\alpha) / \Nus_\omega(\alpha=0) \right)$; the other studies use dimensionless torque $G$ \citep{vanGils2013}, friction coefficient $c_f$ \citep{vandenBerg2005}, or plain torque $\tau$ \citep{Verschoof2016} to define DR.
}
(b) Zoom of the bottom part of (a) where the data from the present work is compared to bubbly drag reduction data using \SI{6}{ppm} of surfactant from \cite{Verschoof2016}}
\label{fig:dragReductionAlpha}
\end{figure}

For a better comparison with bubbly drag reduction, we plot the drag reduction as a function of \red{(gas or particle} volume fraction $\alpha$; see figure \ref{fig:alphaDrAll}. Different studies are shown using different symbols, and $\rey$ is indicated by colours. None of the datasets were compensated for the changes in effective viscosity.
\red{ 
DR is defined in slightly differently way in each study: \cite{vandenBerg2005} makes use of the friction coefficient $\left(1 - c_f(\alpha) /c_f(0) \right)$; \cite{vanGils2013} uses the dimensionless torque 
$G=\tau / (2 \pi L_{mid} \rho \nu^2)$,
$\left(1 - G(\alpha) / G(0) \right)$; and \cite{Verschoof2016} uses the plain torque value $\left(1 - \tau(\alpha) /\tau(0) \right)$.}
While the rigid particles only showed marginal drag reduction, some studies using bubbles achieve dramatic reduction of up to \SI{30}{\percent} and beyond. Figure \ref{fig:alphaDrZoom} shows a zoomed in view of the bottom part of the plot with the rigid sphere data. The triangles denote the data from \cite{Verschoof2016}, corresponding to small bubbles in the Taylor-Couette system. The rigid particles and the small bubbles show a similar drag response. What is remarkable is that this occurs despite the huge difference in size. The estimated diameter of the bubbles in \cite{Verschoof2016} is \SI{0.1}{\milli \metre}, while the rigid spheres are about two orders in magnitude larger. This provides key evidence that the particle size alone is not enough to cause drag reduction\red{, also the density ratio of the particle and the carrier fluid is of importance.
}

\subsection{Effect of marginal changes in particle density ratio}
\begin{figure}
\centering
\subfloat{
  \includegraphics{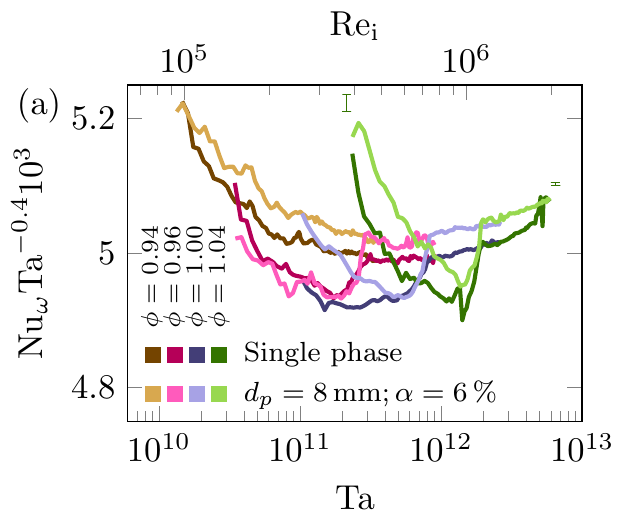}
  \label{fig:TaNuwTaDen}
}
\subfloat{
  \includegraphics{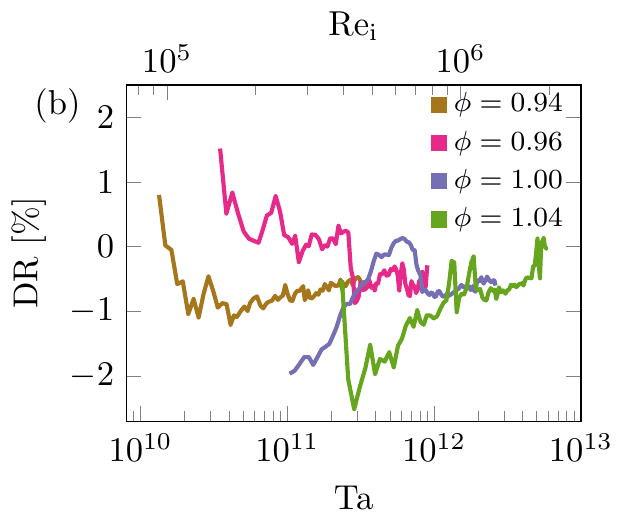}
  \label{fig:TaDrDen}
}
\caption{
(a) $\Tay$ as function of $\Nus_\omega$ compensated by \red{$\Tay^{0.4}$} for various density ratios $\phi=\rho_p/\rho_f$ indicated by the corresponding colour. \red{The darker shades indicate the single phase cases while the lighter shades show the cases using \SI{6}{\percent} particle volume fraction of \SI{8}{\milli\metre} diameter particles.} Due to the increase in viscosity the maximum attainable $\Tay$ is lower for larger density ratios.
\red{
The uncertainty is again estimated using the maximum deviation from the average for multiple runs and here only shown for the green curves. This value is slightly below \SI{1}{\percent} at lower $\Tay$ and decreases with increasing $\Tay$ to values below \SI{0.25}{\percent}. This trend is seen for all $\phi$.
} (b) \red{%
Drag reduction, calculated from the data of figure 5a, plotted against $\Tay$. The drag reduction is defined as $\text{DR} = \left(1 - \Nus_\omega(\alpha=6\%) / \Nus_\omega(\alpha=0) \right)$.
}}
\label{fig:dragReductionDen}
\end{figure}

With the effects of particle size and volume fraction revealed, we next address the sensitivity of the drag to marginal variations in particle density. A change in the particle density ratio brings about a change in the buoyancy and  centrifugal forces on the particle, both of which can affect the particle distribution within the flow. We tune the particle to fluid density ratio $\phi \equiv \rho_p/\rho_f$ by changing the volume fraction of glycerol in the fluid, such that the particles are marginally buoyant ($\phi = 0.94, 0.97$), neutrally buoyant ($\phi=1.00$) and marginally heavy ($\phi=1.04$) particles. In figure \ref{fig:TaNuwTaDen} we show the compensated $\Nus_\omega$ as function of $\Tay$ for various $\phi$. $\alpha$ was fixed to \SI{6}{\percent} and only \SI{8}{\milli\metre} particles were used. The darker shades of colour correspond to the single phase cases, while lighter shades correspond to particle-laden cases. In general, the single phase drag is larger as compared to the particle-laden cases. However, there is no striking difference between the different $\phi$.
In figure \ref{fig:TaDrDen}, we present the drag reduction for particle-laden cases at different density ratios. On average we see for all cases drag modification of approximately $\pm$ \SI{2}{\percent}. We can also identify a small trend in the lower $\Tay$ region: the two larger $\phi$ (heavy and neutrally buoyant particles) tend to have a drag increase, while the smaller $\phi$ cases (both light particles) have a tendency for drag reduction. Nevertheless, the absolute difference in DR between the cases is within 4\%. The above results provide clear evidence that minor density mismatches do not have a serious influence on the global drag of the system.
\red{
To investigate for strong buoyancy effects, additional measurements were done using \SI{2}{\milli\metre} expanded polystyrene particles ($\phi=0.02$).
However, due to the particles accumulating between the inner cylinder segments leading to additional mechanical friction, these measurements were inconclusive.
}

\subsection{Flow statistics using particles} 
\begin{figure}
\centering%
\subfloat{%
  \includegraphics{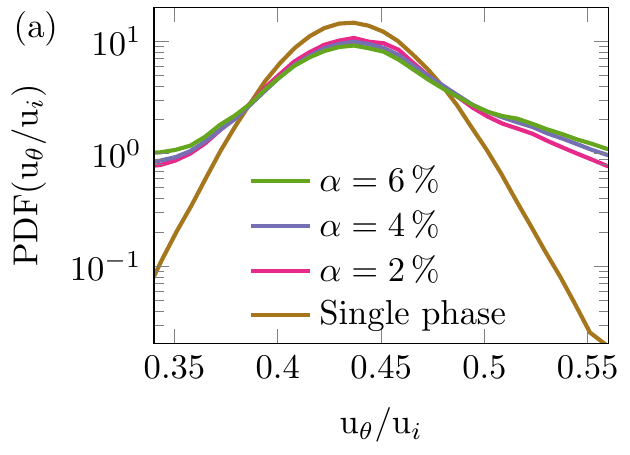}%
  \label{fig:U_w_pdf}
}%
\subfloat{%
  \includegraphics{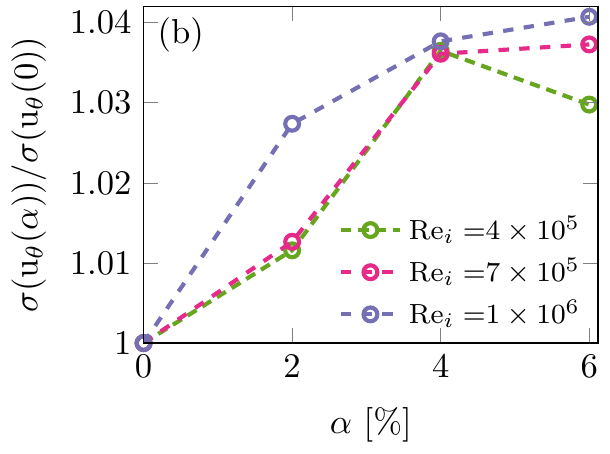}%
  \label{fig:alpha_u_rms_u0}
}%
\caption{%
(a) \red{PDF of $\text{u}_\theta / \text{u}_i$} for various $\alpha$ and the single phase case. The particle size was fixed to \SI{8}{\milli\metre} \red{and $\rey_i=\num{1e6}$} for all cases. (b) Standard deviation of the azimuthal velocity normalized by the standard deviation of the single phase case for three different $\rey$ for a fixed particle size of \SI{8}{\milli\metre}.} \label{fig:ldaresults}
\end{figure}

In the above sections, we presented the effects of changing particle size, volume fraction, and density on the global drag of the system. Next we look into local flow
\red{
properties using LDA while the particles are present.
First, we collected a total of \num{1e6} data points of azimuthal velocity at mid-height and mid-gap.
These were captured over a period of approximately $\num{3e4}$ cylinder rotations. From this data we calculate the probability density function (PDF) of $\text{u}_\theta$ normalised by $u_i$ for various $\alpha$, shown in figure \ref{fig:U_w_pdf}. The particle size was fixed to \SI{8}{\milli\metre} and the Reynolds
number was set to $\num{1e6}$. From this figure we see a large increase in turbulent fluctuations, resulting in very wide tails. While the difference between \SI{2}{\percent}, \SI{4}{\percent}, and \SI{6}{\percent} is not large, we can identify an increase in fluctuations with increasing $\alpha$. These increased fluctuations can be explained by the additional wakes produced by the particles \citep{Poelma2007,Almeras2017}. The increase in fluctuations can also be visualized using the standard deviation of $\sigma(u_\theta) = \left< \text{u}_\theta'^2 \right>^{1/2}$ normalised by the standard deviation of the single phase case---see figure \ref{fig:alpha_u_rms_u0}. In this figure, $\sigma(u_\theta)$ is shown for three different $\rey$, again for \SI{8}{\milli\metre} particles. In general, we see a monotonically increasing trend with $\alpha$, and it seems to approach an asymptotic value. One can speculate that there has to be an upper limit for fluctuations which originate from wakes of the particles. For large $\alpha$ the wakes from particles will interact with each other and with the carried flow.}

Measurements using \SI{4}{\milli\metre} particles yielded qualitatively similar results. It is known that in particle-laden gaseous pipe flows, large particles can increase the turbulent fluctuations, while small particles result in turbulence attenuation \citep{Tsuji1984,Gore1989,Vreman2015}. The LDA measurements were not possible with the smallest particles (\SI{1.5}{\milli\metre}), as the large amount of particles in the flow blocked the optical paths of the laser beams.

\red{
We are confident that for these bi-disperse particle-laden LDA measurements, the large particles do not have an influence on the measurements as these \emph{millimetric}-sized particles are much larger than the fringe spacing ($d_f = \SI{3.4}{\micro\metre}$) and do not show a Doppler burst.
However, during the measurements the particles get damaged and small bits of material are fragmented off the particles. We estimate the size of these particles slightly larger than the tracer particles and these can have an influence on the LDA measurements as they do not act as tracers.
}

\begin{figure}
\centering%
\subfloat{%
  \includegraphics{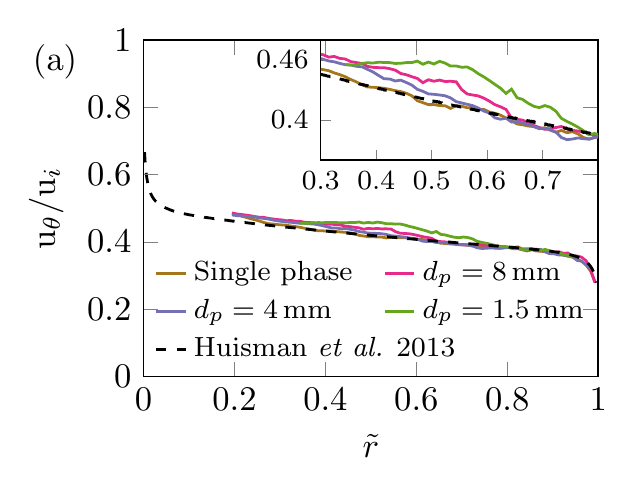}%
  \label{fig:U_theta_radius_size}
}%
\subfloat{%
  \includegraphics{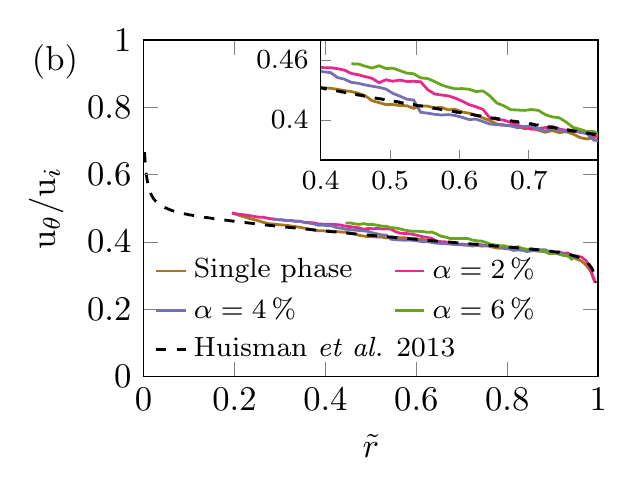}%
  \label{fig:U_theta_radius_vol}
}%
\caption{%
\red{
$\text{u}_\theta$ normalised by the velocity of the inner cylinder wall $\text{u}_i$ as function of the normalised radius for various $d_p$ while $\alpha=\SI{2}{\percent}$ (a) and various $\alpha$ while $d_p=\SI{8}{\milli\metre}$ (b). In all cases $\rey_i$ is fixed to $\num{1e6}$. For comparison, the single phase case using water at $\rey_i=\num{1e6}$ from \cite{Huisman2013} is also plotted in dashed black in both plots. Both figures have an inset showing an enlargement of the centre area from the same figure.}} \label{fig:ldaradial}
\end{figure}

\red{
How the average azimuthal velocity changes with particle radius is shown in figure \ref{fig:ldaradial}.
We measured a total of $\num{3e4}$ data points during approximately 900 cylinder rotations. Again, the data were corrected for velocity bias by using the transit time as a weighing factor. Figure \ref{fig:U_theta_radius_size} shows the effect of particle size for $\alpha=\SI{2}{\percent}$, and figure \ref{fig:U_theta_radius_vol} shows the effect of particle volume fraction for \SI{8}{\milli\metre} particles. Both figures additionally show the high-precision single phase data from \cite{Huisman2013} for which our single phase measurements are practically overlapping. Since LDA measurements close to the inner cylinder are difficult, due to the reflecting inner cylinder surface, we limited our radial extent to $\tilde{r}=(r - r_i) / (r_o - r_i)=[0.2,1]$. We found that the penetration depth of our LDA measurements is the smallest for experiments with the smallest particles and the largest $\alpha$. All differences with the single phase case are only marginal and we can conclude that the average mean velocity is not much affected by the particles in the flow, at least for $\tilde{r} \geq 0.2$.}

\begin{figure}
\centering%
\includegraphics{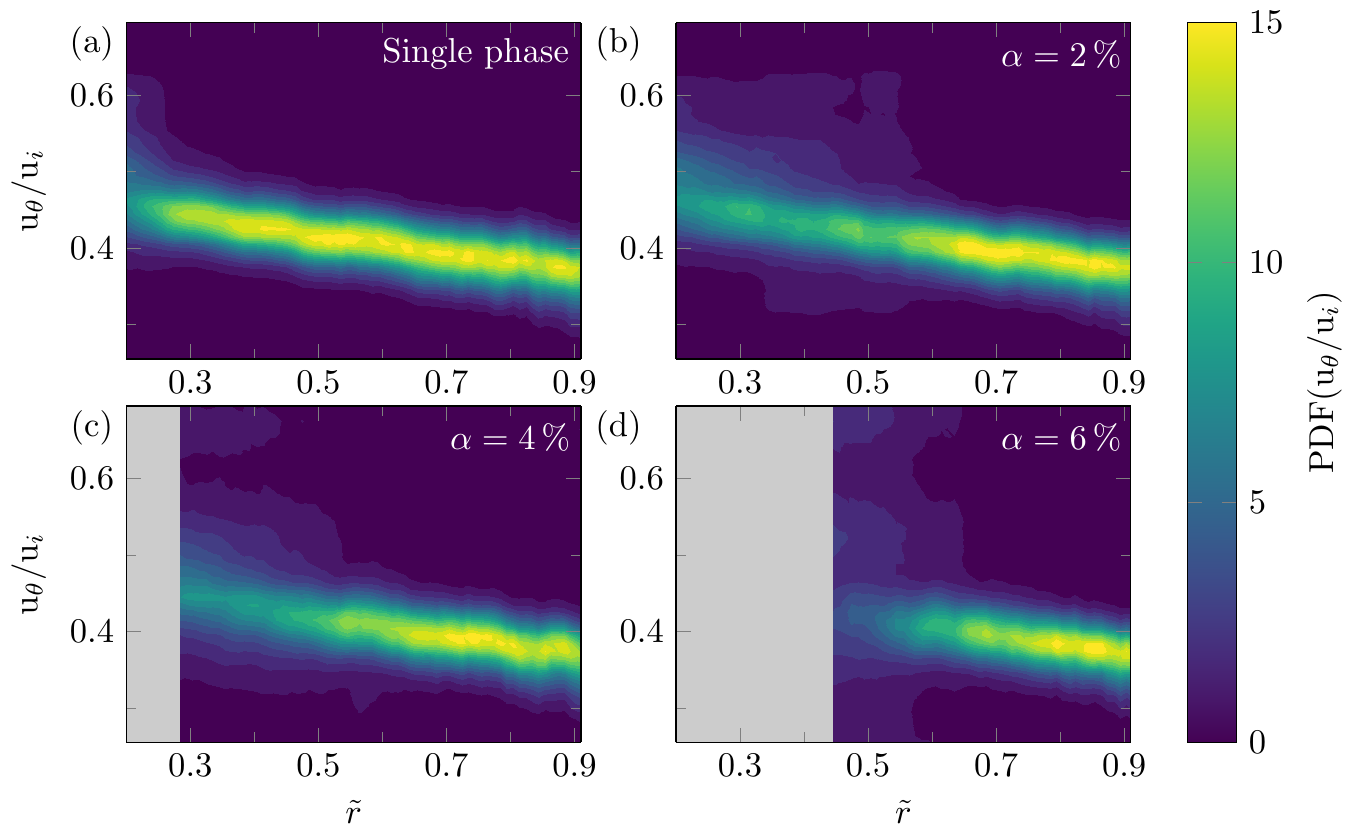}%
\caption{%
\label{fig:heatmap}
\red{
PDF of the normalised azimuthal velocity as a function of normalised radial position for various $\alpha$ for the case of \SI{8}{\milli\metre} particles and the single phase case while keeping $\rey$ at $\num{1e6}$. With increasing $\alpha$ the maximum penetration depth decreases. The grey areas indicate radial positions for which no data is available.
}}
\end{figure}

\red{
To get an idea of the fluctuations we can use the previous data to construct a two-dimensional PDF of the azimuthal velocity as function of radius. These are shown for $\rey = \num{1e6}$ using \SI{8}{\milli\metre} particles at various $\alpha$ and the single phase case in figure \ref{fig:heatmap}.
First thing to notice is again that the penetration depth is decreasing with increasing $\alpha$.
The single phase case shows a narrow banded PDF. When $\alpha$ is increased, for the lower values of $\tilde{r}$ the PDF is much wider.
While it makes sense that an increase in $\alpha$ increases the fluctuations due to the increased number of wakes of particles, this is expected everywhere in the flow, not only closer to the inner cylinder.
It is possible that the particles have a preferred concentration closer to the inner cylinder. 
We have tried to measure the local concentration of particles as function of radius but failed due to limited optical accessibility.
Therefore, we can only speculate under what circumstances there would be an inhomogeneous particle distribution which would lead to the visible increase in fluctuations.
The first possibility is a mismatch in density between the particle and the fluid, which would result in light particles ($\phi < 1$) to accumulate closer to the inner cylinder.
Another possibility is that due to the rotation of the particle, an effective lift force arises, leading to a different particle distribution in the flow. While this is quite plausible, this is difficult to validate as we would need to capture the rotation.
The fragments of plastic that are sheared off the particles can also have a bias to the LDA measurement. While we estimate them to be larger than the tracers, they might still be small enough to produce a signal and they might not follow the flow faithfully.
}

\section{Conclusions and outlook}\label{sec:conclusions_and_outlook}

We have conducted an experimental study on the drag response of a highly turbulent Taylor-Couette flow containing rigid neutrally buoyant spherical particles. We have found that, unlike the case of bubbles used in prior works~\citep{vanGils2013,Verschoof2016}, rigid particles barely reduce (or increase) the drag on the system, even for cases where their size was comparable to that of bubbles used in other studies.
\red{
There was no significant size effect. Even for very large particles, which can attenuate turbulent fluctuations and generate wakes, there was no distinct difference with the single phase flow.
}
We also varied the volume fraction of the particles in the range 0\%--6\%. The particle volume fraction has no greater effect on the system drag than what is expected due to changes in the apparent viscosity of the suspension. Further, we tested the sensitivity of our drag measurements to marginal variations in particle to fluid density ratio $\phi$. A trend was noticeable, towards drag reduction when $\phi$ was reduced from 1.00 to 0.94. This suggests that a low density of the particle could be a necessary ingredient for drag reduction.
Finally, we have also probed the local flow at the mid height and mid gap of the system using LDA. With the addition of particles, the liquid velocity fluctuations are enhanced, with wider tails of the distributions. A finite relative velocity between the particle and the flow around it can cause this increase in velocity fluctuations~\citep{Mathai2015}, as seen for bubbly flows (pseudo-turbulence), and in situations of sedimenting particles in quiescent or turbulent environments~\citep{Gore1989}.  In the present situation, the relative velocity between the particle and the flow is expected, owing to the inertia of the finite-sized particles we used.
\red{
There is only a marginal deviation from the single phase case in the average azimuthal velocity over the radial positions measured using any size or concentration of particles measured. From the two-dimensional PDFs, we see that closer to the inner cylinder, using smaller $d_f$ or larger $\alpha$, the PDF gets wider. This can be due to a preferential concentration of the particles or a slight density mismatch.
}

\red{Our study is a step towards a better understanding of the mechanisms of bubbly drag reduction. Bubbles are deformable, and they have a tendency to migrate towards the walls, either due to lift force~\citep{Dabiri2013}, or due to the centripetal effects~\citep{vanGils2013}. When compared to the drag reducing bubbles in \cite{vanGils2013,Verschoof2016}, our particles do not deform, and they do not \red{experience centripetal effects as they are density matched}.
At least one of these differences must therefore be crucial for the observed, bubbly drag reduction in those experiments.
\red{
In a future investigation, we will conduct more experiments using very light spherical particles that experience similar centripetal forces as the bubbles in \cite{vanGils2013}, but are non-deformable.}
These particle need to be larger than the size of the gap between the inner cylinder segments and very rigid, or the setup needs to be modified to close the gap between the IC segments.
Such experiments can then disentangle the role of particle density on drag reduction from that of the particle shape.
}

\emph{Acknowledgements}: We would like to thank Elisabeth Guazzelli, Bert Vreman, Rodrigo Ezeta, Pim Bullee and Arne te Nijenhuis for various stimulating discussions. Also, we like to thank Gert-Wim Bruggert and Martin Bos for technical support. This work was funded by STW, FOM, and MCEC, which are part of the Netherlands Organisation for Scientific Research (NWO). CS acknowledges financial support from VIDI grant No. 13477, and the Natural Science Foundation of China under Grant No. 11672156.

\bibliographystyle{jfm}

\bibliography{JFM_rigidParticles}

\end{document}